\documentclass[aps,prd,twocolumn,floats,floatfix,showpacs,superscriptaddress,nofootinbib,preprintnumbers]{revtex4-1}

\usepackage{graphicx,amssymb,amsmath,amsthm,amsfonts,epsfig,epsf,fixmath}
\usepackage[usenames]{color}
\usepackage{epstopdf}

\usepackage{aas_macros}
\usepackage{bm}
\usepackage{dcolumn}
\usepackage[dvipsnames]{xcolor}
\usepackage[colorlinks]{hyperref}
\usepackage{cleveref}
\usepackage[utf8]{inputenc}
\usepackage{resizegather}

\usepackage{multirow}
\usepackage{cleveref}

\newcommand{\GSSI}{Gran Sasso Science Institute (GSSI), I-67100 L’Aquila, Italy}
\newcommand{\GranSasso}{INFN, Laboratori Nazionali del Gran Sasso, I-67100 Assergi, Italy}

\begin{document}

\title{Impact and detectability of spin-tidal couplings in neutron star inspirals}

\author{Gon\c{c}alo Castro}
	\email{goncalo.castro@uniroma1.it}
	\affiliation{Dipartimento di Fisica, ``Sapienza" Università di Roma \& Sezione INFN Roma1, Piazzale Aldo Moro 
5, 00185, Roma, Italy}

\author{Leonardo Gualtieri}
	\affiliation{Dipartimento di Fisica, ``Sapienza" Università di Roma \& Sezione INFN Roma1, Piazzale Aldo Moro 
5, 00185, Roma, Italy}

 \author{Andrea Maselli}
\address{\GSSI}
\address{\GranSasso}

\author{Paolo Pani}
	\affiliation{Dipartimento di Fisica, ``Sapienza" Università di Roma \& Sezione INFN Roma1, Piazzale Aldo Moro 
5, 00185, Roma, Italy}

\begin{abstract}
The gravitational wave signal from a binary neutron star merger carries the imprint of the deformability properties of the coalescing bodies, and then of the equation of state of neutron stars. In current models of the waveforms emitted in these events, the contribution of tidal deformation is encoded in a set of parameters, the tidal Love numbers. 
More refined models include tidal-rotation couplings, described by an additional set of parameters, the rotational tidal Love numbers, which appear in the waveform at $6.5$ post-Newtonian order.
For  neutron stars with spins as large as $\sim0.1$, we show that neglecting tidal-rotation couplings may lead to a significant error in the parameter estimation by third-generation gravitational wave detectors.
By performing a Fisher matrix analysis we assess the measurability of rotational tidal Love numbers, showing that their contribution in the waveform could be measured by third-generation detectors.  Our results suggest that current models of tidal deformation in late inspiral should be improved in order to avoid waveform systematics and extract reliable information from gravitational wave signals observed by next generation detectors.
\end{abstract}

\preprint{ET-0072A-22}
\maketitle

\section{Introduction and summary} \label{sec:intro}


The gravitational-wave~(GW) signal emitted during the latest stages of a neutron-star~(NS) coalescence before the merger  significantly depends on how a NS gets deformed by the gravitational field of its companion. This effect is quantified by a set of parameters called tidal Love numbers~(TLNs)~\cite{Murraybook,PoissonWill}, which encode the tidal deformability properties of the star, and depend on the NS equation of state~(EoS)~\cite{Flanagan:2007ix,Hinderer:2007mb}.
Extracting the TLNs from GW detections provides a way to measure the NS EoS~\cite{Baiotti:2010xh,Baiotti:2011am,Vines:2011ud,Pannarale:2011pk,Vines:2010ca,Lackey:2011vz,Lackey:2013axa,Favata:2013rwa,Yagi:2013baa,Maselli:2013mva,Maselli:2013rza,DelPozzo:2013ala,TheLIGOScientific:2017qsa,Bauswein:2017vtn,Most:2018hfd,Harry:2018hke,Annala:2017llu,EoS-GW170817,Akcay:2018yyh,Maselli:2020uol} (see Refs.~\cite{GuerraChaves:2019foa,Chatziioannou:2020pqz} for some reviews), to constrain alternative theories of gravity in an EoS-independent fashion~\cite{Yagi:2013awa,Yagi:2013bca} (see Ref.~\cite{Yagi:2016bkt} for a review), and to test the nature of dark compact objects other than black holes~\cite{Cardoso:2017cfl} (see Refs.~\cite{Cardoso:2019rvt,Maggio:2021ans} for some reviews).

As current GW detectors approach design sensitivity, it is reasonable to expect that the GW event catalogue~\cite{LIGOScientific:2021djp} will be enlarged with several binary NSs and mixed black hole-NS binaries, possibly with higher signal-to-noise ratio~(SNR) than the prototypical signal GW170817~\cite{TheLIGOScientific:2017qsa}. This will certainly be the case in the era of third-generation~(3G) GW detectors~\cite{Kalogera:2021bya}, such as Cosmic Explorer~\cite{Reitze:2019iox} and the Einstein Telescope~(ET)~\cite{Hild:2010id,Maggiore:2019uih}.
Indeed, even just a single detection of a binary NS coalescence by a 3G detector will constrain the properties of nuclear matter, through the measurements of the TLNs, to unprecedented levels~\cite{Pacilio:2021jmq}. 

However, the high SNR expected in the 3G era also urges to reduce possible waveform systematics~\cite{Narikawa:2019xng,Gamba:2020wgg} by improving current waveform models, including all possible effects related to the tidal deformability of NSs. 
For this reason, current models of the late inspiral of coalescing NS binaries, in which the effect of tidal deformation is described in terms of a single parameter~\cite{Dietrich:2017aum,Dietrich:2018uni} --~namely a combination of the quadrupolar electric TLNs of the two bodies~-- should be extended to account for higher post-Newtonian~(PN) effects. The latter possibly include higher-order and magnetic TLNs~\cite{Landry:2015cva,Landry:2015snx,Abdelsalhin:2018reg,Pani:2018inf,Banihashemi:2018xfb,Henry:2020ski}, the so-called rotational TLNs~(RTLNs), which arise from the coupling between the object's angular momentum and the external tidal field~\cite{Pani:2015nua,Landry:2015zfa,Landry:2017piv,Gagnon-Bischoff:2017tnz,Abdelsalhin:2018reg,Jimenez-Forteza:2018buh,Castro:2021wyc}, as well as the effects of time-dependence of the tidal field and dynamical tides~\cite{Lai:1993di,Reisenegger:1994,Ho:1998hq,Landry:2015cva,Hinderer:2016eia,Steinhoff:2016rfi,Poisson:2020mdi,Poisson:2020vap,Steinhoff:2021dsn}.

In this article we quantify the impact of spin-tidal couplings, and in particular of the RTLNs, in the parameter estimation from a binary NS waveform. We extend the analysis of~\cite{Jimenez-Forteza:2018buh} in two main directions: i)~Using the recent computation of the RTLNs for static 
fluids and the associated hidden symmetry unveiled in~\cite{Abdelsalhin:2018reg,Castro:2021wyc}, we employ an inspiral waveform approximant that coherently includes \emph{all} static tidal effects up to $6.5$-PN order. ii)~We perform a statistical analysis based on the Fisher-information matrix (FIM), which accounts for correlations among the parameters.

The rest of the paper is organized as follows. In Sec.~\ref{sec:tidint} we summarize the theory of tidal deformations of rotating compact stars, introducing the TLNs and RTLNs, and the PN waveform up to $6.5$ order. In Sec.~\ref{sec:stat} we briefly describe the statistical analysis based on the FIM. In Sec.~\ref{sec:results} we discuss our results on the impact of spin-tidal couplings on the waveform, and on the measurability of the RTLNs. Finally, in Sec.~\ref{sec:conclusions} we draw our conclusions.

\section{Tidal interaction of rotating compact stars}\label{sec:tidint}
Here we briefly discuss the tidal deformations of rotating compact bodies, describing the TLNs and the RTLNs and showing how the coupling between rotation and tidal interaction affects the gravitational waveform emitted by a binary NS. We use units with $G=1$. Greek letters denote spacetime indices ($\mu,\nu=0,\dots,3$), while Latin letters denote space indices ($a,b=1,\dots,3$). We shall follow the notation of~\cite{Abdelsalhin:2018reg} and~\cite{Jimenez-Forteza:2018buh}.
\subsection{Tidal deformations of rotating compact stars}
The general relativistic theory of tidal deformations of nonrotating compact objects  has been developed in~\cite{Flanagan:2007ix,Hinderer:2007mb,Binnington:2009bb,Damour:2009vw}, where it was shown that when a static, spherically symmetric compact star is perturbed by a (static) external tidal field, it acquires mass and current  multipole moments ($Q^{a_1\cdots a_l}$, $S^{a_1\cdots a_l}$, respectively~\cite{Geroch:1970cd,Hansen:1974zz}) given by:
\begin{align}
    Q^{a_1\cdots a_l}&=\lambda_lG^{a_1\cdots a_l}\nonumber\\
    S^{a_1\cdots a_l}&=\sigma_lH^{a_1\cdots a_l}
    \label{eq:adiab_nonrot}
\end{align}
where $l\ge2$, $G^{a_1\cdots a_l}$, $H^{a_1\cdots a_l}$ are the {\it electric} and {\it magnetic} components of the tidal field, and $\lambda_l$, $\sigma_l$ are the electric and magnetic TLNs, respectively. The TLNs of a NS depend on the EoS of the star and on its mass, and can be computed using perturbation theory~\cite{Hinder:2008kv}.
The two above equations are also called {\it adiabatic relations}, because they can be applied to the inspiral of a binary NS, when each star is tidally deformed by the companion, as long as the adiabatic approximation is satisfied, i.e. the tidal field is approximately constant over the time scale of the stellar response.

If rotation is included  in the model~\cite{Poisson:2014gka,Pani:2015hfa,Pani:2015nua,Landry:2015zfa,Poisson:2016wtv}, it introduces couplings between tidal field components and multipole  moments having different parities (electric vs. magnetic) and with different values of $l$. Neglecting the contributions with $l>3$, the adiabatic relations~\eqref{eq:adiab_nonrot} generalize to
\begin{align}
Q^{ab} = & \lambda_{2} G^{ab} + \frac{\lambda_{23}}{c^2} J^c H^{abc} \nonumber\\
Q^{abc} = & \lambda_{3} G^{abc} + \frac{\lambda_{32}}{c^2} J^{\langle c} H^{ab \rangle} \nonumber\\
S^{ab} = & \frac{\sigma_{2}}{c^2} H^{ab}  + \sigma_{23} J^c G^{abc} \nonumber \\
S^{abc} = & \frac{\sigma_{3}}{c^2} H^{abc} + \sigma_{32} J^{\langle c} G^{ab \rangle}
\label{eq:adiabatic}
\end{align} 
where $J^a$ is the spin vector,  $\lambda_{ll'}$, $\sigma_{ll'}$ are the RTLNs with electric and magnetic parity, respectively, 
and $\langle\cdots\rangle$ denotes trace-free symmetrization.

In the above derivation the spacetime is assumed to be stationary, $g_{\mu\nu,t}=0$, while the fluid four-velocity can only have a nonvanishing azimuthal component $u^\varphi$, where $\varphi$ is the azimuthal angle associated to rotation. In~\cite{Pani:2015nua} it was also assumed that the fluid perturbations induced by the tidal field are static, i.e. that $\delta u^\varphi=0$. More recently, is was noted that in actual binary NS systems, the stationary limit of a time-dependent compact star has arguably {\it irrotational} perturbations~\cite{Landry:2015cva,Pani:2018inf}, in which $\delta u^\varphi$ is determined by imposing the vanishing of the vorticity tensor (see also~\cite{Castro:2021wyc} for further details). 

Like the TLNs, also the RTLNs depend on the mass of the star and on its EoS, and can be determined using perturbation theory. This computation is rather involved: it requires solving a large system of coupled ordinary differential equations, describing the gravitational and fluid perturbations with different parities and different values of the harmonic index $l$. A preliminary computation in~\cite{Pani:2015nua} turned out to be affected by some errors in the numerical implementation; finally, in~\cite{Castro:2021wyc} we have computed the RTLNs associated to static perturbations of NSs. 
The explicit computation of~\cite{Castro:2021wyc} also allowed us to confirm the existence of a ``hidden symmetry'' --~which had been first proposed in~\cite{Abdelsalhin:2018reg}~-- between (static) electric and magnetic RTLNs:
\begin{equation}
    \sigma_{32}=4\lambda_{23}\,,\qquad\lambda_{32}=2\sigma_{23}\label{eq:hiddens}\,.
\end{equation}
The above relations effectively halve the number of RTLNs that should be computed to fully characterize the tidal contribution in a NS waveform model, as summarized below.

\subsection{Gravitational waveform from tidally deformed compact binaries}
Let $M_1$, $M_2$ be the masses of the two compact bodies in circular orbit, and $J_1$, $J_2$ their angular momenta. Be $M=M_1+M_2$, $\eta_A=M_A/M$ ($A=1,2$), $\chi_A=c J_A/M_A^2$ their dimensionless spin parameters, $\lambda_l^{(A)}$, $\sigma_l^{(A)}$, $\lambda_{ll'}^{(A)}$, $\sigma_{ll'}^{(A)}$ their TLNs and RTLNs. Moreover, let $\nu=\eta_1\eta_2$ be the symmetric mass ratio, $\omega$ the orbital angular velocity of the binary, and $x=(M\omega)^{2/3}/c^2$. 

We model the GW signal emitted by the compact binary system using the TaylorF2 approximant in the frequency domain~\cite{Arun:2008kb,Buonanno:2009zt,Mishra:2016whh}. The GW phase can be written as the sum of a point-particle contribution and of a tidal term, $\psi(x)=\psi_{\rm pp}(x)+\psi_{\rm T}(x)$. 
The former depends on the mass and spin components and includes up to $\mathcal{O}(x^{7/2})$, namely $3.5$-PN, corrections. For brevity, we show here its form up to $1.5$-PN order:
\begin{align}
    \psi_{\rm pp}(x)=&\frac{3}{128\nu x^{5/2}}
    \left\{1+\left(\frac{3715}{756}+\frac{55}{9}\nu\right)x\right.\nonumber\\
    &+\left(\frac{113}{3}\left(\eta_1\chi_1+\eta_2\chi_2\right)-\frac{38}{3}\nu\left(\chi_1+\chi_2\right)\right.\nonumber\\
    &\left.\left.-16\pi\right)x^{3/2}+O(x^2)\right\}\,.
    \label{eq:psiPP}
\end{align}
The explicit expression of $\psi_{\rm pp}(x)$ to $3.5$-PN order can be found in Appendix~\ref{app:Taylor}.

The leading-order contribution describing tidal interactions, $\psi_{\rm T}$, enters at the $5$-PN order\,\footnote{Note that the tidal contribution to the GW phase is much larger than a naive counting of their PN order could suggest,  being magnified by the dimensionless quantity $(c^2 R/M)^5$ that appears in the TLNs.}, and depends  on a linear combination of the quadrupolar, electric TLNs $\lambda_2^{(A)}$ of the two bodies. Electric TLNs with $l>2$ and magnetic TLNs contribute to higher PN order in the waveform. We consider the tidal phase including up to $\mathcal{O}(x^{6.5})$~\cite{Abdelsalhin:2018reg,Abdelsalhin:2019ryu}
\begin{align}
 \psi_{\rm T}(x) &= \frac{3}{128 \nu x^{5/2}} \left\{ -\frac{39}{2}\tilde\Lambda x^5\right. \nonumber \\
 & \left. + \left(-\frac{3115}{64}\tilde{\Lambda} + (\eta_1-\eta_2)\frac{6595}{364}\delta\tilde{\Lambda} + \tilde{\Sigma}\right) x^6\right.\nonumber\\
 & \left. + (\hat\Lambda+\hat\Sigma+ \hat\Gamma+ \hat K ) x^{6.5} +{\cal O}(x^7)\right\}\,, \label{eq:psiT}
\end{align}
where $\Lambda_A=\lambda_2^{(A)}/M_A^5$, $\Sigma_A=\sigma_2^{(A)}/M_A^5$ are the dimensionless TLNs,
\begin{align}
 \tilde\Lambda &=\frac{16}{13}\left(\frac{12}{\eta_1}-11\right) \eta_1^5\Lambda_1+  (1\leftrightarrow 2)\,,\label{eq:defLambda}\\
 \delta\tilde{\Lambda} & = \frac{1}{2}\left[(\eta_1-\eta_2)\left(1-\frac{13272}{1319}\nu + \frac{8944}{1319}\nu^2\right)(\Lambda_1+\Lambda_2) + \right.\nonumber\\
 & \left. \left(1 - \frac{15910}{1319}\nu + \frac{32850}{1319}\nu^2 + \frac{3380}{1319}\nu^3\right)(\Lambda_1-\Lambda_2)\right]\label{eq:deltaLambda}\\
 \tilde\Sigma &= 
  \left( \frac{6920}{7} - \frac{20740}{21 \eta_1} \right)   \eta_1^5\Sigma_1   +(1\leftrightarrow 2)\,,
\end{align}
\begin{align}
\hat\Lambda &=
  \left[ \left( \frac{593}{4} - \frac{1105}{8 \eta_1}+\frac{567 \eta_1}{8} -81 \eta_1^2 \right) \chi_2\right. \nonumber\\
  &+\left.\left( -\frac{6607}{8} +\frac{6639 \eta_1}{8} -81 \eta_1^2 \right) \chi_1 \right] \eta_1^5\Lambda_1\nonumber\\
  &+(1\leftrightarrow 2)\,, \label{eq:hatLambdaE}\\
 \hat\Sigma &=\left[\left(-\frac{9865}{3} + \frac{4933}{3 \eta_1} + 1644 \eta_1 \right) \chi_2 -\chi_1 \right] \eta_1^5\Sigma_1\nonumber \\
 &+(1\leftrightarrow 2)\,, \label{eq:hatLambdaM} \\
 \hat K &=\frac{39}{2}\pi \tilde\Lambda\,, \label{eq:hatK} \\
\hat\Gamma &=\frac{\chi_1}{M^4}\left[ \left(  856 \eta_1 - 816 \eta_1^2 \right){\lambda_{23}^{(1)}} \right.\nonumber\\
&\left.- \left(\frac{833 \eta_1}{3} - 278 \eta_1^2 \right){\sigma_{23}^{(1)}}\right. \nonumber\\
& \left. - \nu \left(272 {\lambda_{32}^{(1)}} -204 {\sigma_{32}^{(1)} }\right)\right]+(1\leftrightarrow 2) \,.
\label{eq:hatGamma}
 \end{align}
For the $6$-PN term in Eq.~\eqref{eq:psiT} we follow the conventions of~\cite{Lackey:2015}, which splits the quadrupolar, electric tidal parameters into $\tilde{\Lambda}$ and $\delta\tilde{\Lambda}$. This choice improves the measurability of the tidal deformability because $\tilde\Lambda$ appears both in the $5$-PN and in the $6$-PN terms. Note that $\delta\tilde\Lambda$ identically vanishes for equal-mass binaries.


The spin-tidal couplings appear in the GW phase through the $6.5$-PN terms $\hat\Lambda$, $\hat\Sigma$, $\hat\Gamma$.
While $\hat\Lambda$, $\hat\Sigma$ (also studied in~\cite{Jimenez-Forteza:2018buh}) depend on the TLNs $\lambda_2$, $\sigma_2$, $\hat\Gamma$ is proportional to the  RTLNs $\lambda_{23}$, $\sigma_{23}$, $\lambda_{32}$, $\sigma_{32}$ of the two bodies. Due to the hidden symmetry~\eqref{eq:hiddens}, this term only depends on {\it two} independent RTLNs (for each object). 

We remark that Eq.~\eqref{eq:hatGamma} builds on the assumption that fluid perturbations are static. The contribution of the RTLNs to the tidal GW phase has not been derived in the case of irrotational perturbations, since the computation is much more involved than for static perturbations\,\footnote{As argued in~\cite{Castro:2021wyc}, the irrotational case seems to require the derivation of the field equations starting from a time-dependent interaction Lagrangian.}. 
Arguably, the amplitude of such effect is comparable with the one considered in this work, and thus our analysis provides a reliable order-of-magnitude estimate of the impact on the GW signal of irrotational RTLNs as well.

\section{Statistical analysis}\label{sec:stat}
The output in time $d(t)$ of a GW interferometer is given by
\begin{equation}
	d(t) = h(t;\vec\theta,\vec\gamma) + n(t),
\end{equation}
where $h$ is the GW signal, and $n$ is a given realization of the detector noise. The former is fully specified by the intrinsic (or physical) parameters $\vec{\gamma}$, such as the binary masses, spins and Love numbers, and by the extrinsic parameters $\theta$, which define the source distance, sky orientation and polarization with respect to the detector. 
Assuming stationariety, stochasticity and Gaussianity, $n(t)$ can be described in terms of a frequency dependent noise spectral density $S_n(f)$~\cite{Sathyaprakash:2009xs}. 
Assessing the presence of a GW signal within the detector stream requires a proper figure of merit, such as the matched-filter SNR $\rho$, defined as
\begin{equation}
	\rho^2 = \left(d|h_T\right) = \left(h|h_T\right) + \left(n|h_T\right).
\end{equation}
Here, $h_T(\vec{\theta}_T,\vec{\gamma}_T)$ corresponds to a specific waveform of a template bank, identified by the set of intrinsic and extrinsic parameters $(\vec{\gamma}_T,\vec{\theta}_T)$.
The inner product is defined as
\begin{equation}
	(h|h_T) = 4\mathcal{R}\int_{f_{\rm min}}^{f_{\rm max}}df\frac{\tilde{h}(f)\tilde{h}_T^*(f)}{S_n(f)},
\end{equation}
with $f_{\rm min}$ and $f_{\rm max}$ cutoff frequencies, characteristic of each detector.

Once the GW signal has been correctly identified, Bayesian information theory can be applied to determine the posterior probability distribution of the waveform parameters
\begin{equation}
	p(\vec{\gamma}_T|d) \propto p_0(\vec{\gamma}_T) e^{-\frac{1}{2}\left(d-h_T|d-h_T\right)},
	\label{eq:bayes}
\end{equation}
where $p_0(\gamma_T)$ is the prior probability distribution. Under the assumptions of high SNR and flat priors, this expression can be rewritten as~\cite{Chatziioannou:2017tdw}
\begin{equation}
	p(\vec{\gamma}_T) \propto \exp\left[-\rho^2\left(1-\mathcal{M}(h|h_T)\right)\right],
	\label{eq:match_posterior}
\end{equation}
in terms of the match $\mathcal{M}(h|h_T)$ between the signal $h$ and a template waveform $h_T$. The match is defined as a normalized inner product maximized over the waveform extrinsic parameters,
\begin{equation}
	\mathcal{M}(h(\vec{\gamma}_0),h_T(\vec{\gamma}_T)) = \underset{\vec{\theta}_0, \vec{\theta}_T}{\rm max}\frac{\left(h|h_T\right)}{\sqrt{\left(h|h\right)\left(h_T|h_T\right)}},
\end{equation}
which serves as a measure of the metric distance between two waveform representations. Because of this property, Eq.~\eqref{eq:match_posterior} is useful to study the bias induced on the recovered parameters by our choice of template waveform, and therefore to study the possibility of systematic errors.

Equation~\eqref{eq:bayes} also provides the basic ingredient to compute statistical uncertainties on the source parameters. Working again in the limit of large SNR, one can rewrite the posterior distribution as~\cite{Vallisneri:2007ev}
\begin{equation}
	p(\vec{\gamma}_T|d) \propto p_0(\vec{\gamma}_T) e^{-\frac{1}{2}\Gamma_{ij}\Delta\gamma^i\Delta\gamma^j}\,,
	\label{eq:fisher_posterior}
\end{equation}
where $\Delta\gamma^i = \gamma_T^i - \gamma_0^i$ is the deviation of the estimated parameters from their true values, and $\Gamma_{ij} = (\partial_i h|\partial_j h)$ is the FIM, which corresponds to the inverse of the (likelihood) covariance matrix among the waveform parameters.
For flat priors, statistical errors are simply given by 
\begin{equation}
	\sigma_{\gamma^i} = \sqrt{\left(\Gamma^{-1}\right)_{ii}}\ .
	\label{eq:uncertainty}
\end{equation}
As discussed in Sec.~\ref{sec:results} we use general priors, for which  parameter errors are no longer given by the direct inversion of the FIM. In order to properly incorporate our prescription for $p_0(\vec\gamma_T)$, we have devised the following strategy. We have built the semi-analytic posterior distribution \eqref{eq:fisher_posterior} using the FIM to define the likelihood function for $\vec\gamma_T$, and multiplying afterward by the two priors we have imposed on $\chi$ and $\delta\tilde{\Lambda}$. Values from the joint posterior on the source parameters are then obtained by sampling $p(\vec\gamma_T|d)$ with a Monte Carlo Markov Chain algorithm.
We sample the posterior distribution using the {\tt emcee} algorithm with stretch move \cite{Foreman_Mackey_2013}. 
For each set of data, we run 20 walkers of $10^5$ samples.
 This simple procedure avoids in general the direct inversion of $\Gamma_{ij}$, and allows computing the covariance matrix of $\vec\gamma_T$ for any choice of the prior functions\,\footnote{We have checked that our method reproduces the parameter's errors by direct inversion of the FIM, when no priors are imposed.}.

\begin{figure*}[th]
\centering
    \includegraphics[width=\textwidth]{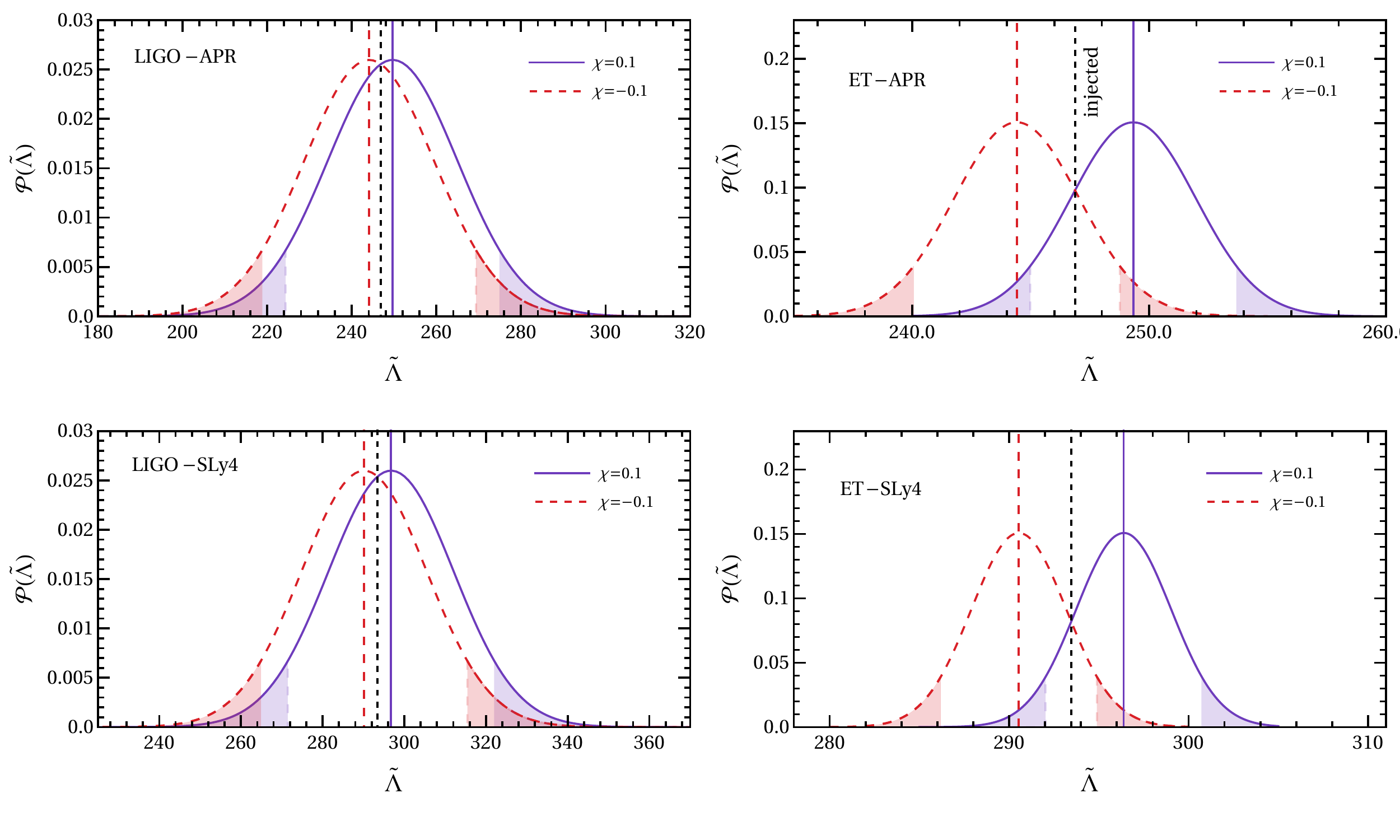}
    \caption{Probability distribution on $\tilde \Lambda$ inferred from the match for Setup~I ($h_\text{spin-tidal}$ vs $h_\text{tail}$), i.e. quantifying the bias introduced by the spin-tidal terms in the waveform. Top and bottom rows refer to APR and SLy4 EoS respectively, while left and right columns correspond to parameter estimation performed for LIGO-Virgo and ET. The solid and dashed vertical lines identify the mean value of each distribution. The vertical short-dashed line in each panel provides the injected value of $\tilde{\Lambda}$. The white area between shaded regions 
    correspond to the $90\%$ probability distribution for 
    ${\cal P}(\tilde{\Lambda})$.
    \label{fig:bias_spintidal}}
\end{figure*}

\section{Results} \label{sec:results}
Based on the methods summarized in Sec.~\ref{sec:stat}, we study both the impact of the spin-tidal couplings on the systematics errors on the tidal deformability, and the measurability of the RTLNs appearing in the $6.5$-PN order tidal term in the GW phase. For simplicity, we assume high SNR and Gaussian noise.

Our reference for this study is the first binary NS event detected by the LIGO-Virgo Collaboration, GW170817~\cite{TheLIGOScientific:2017qsa}, and for simplicity we assume equal masses $M_1=M_2=1.4M_\odot$. We also assume small spins $\chi_1$ and $\chi_2$ throughout. As for the (R)TLNs, we compute them for different EoSs that are realistic in terms of both their predicted maximum masses and tidal deformabilities. We further consider the internal fluid of the stars to be static, and compute the tidal quantities accordingly.
As discussed in Sec.~\ref{sec:tidint}, we expect {\it irrotational} RTLNs to have a comparable impact on the gravitational waveform. In that case, our results should be considered as an order-of-magnitude estimate.

We perform our computation for two possible detectors, LIGO-Virgo and ET. We consider the former for a network of three detectors with the same design sensitivity\cite{LIGO_design}, in a frequency range of $[9,2048]$~Hz, and the latter in its ET-D configuration~\cite{Hild:2010id} assuming a single detector in a triangular configuration\footnote{In practice, we multiply the ET-D sensitivity curve~\cite{Hild:2010id} by a factor $2/\sqrt{3}$ to account for a triangular geometry.} in the frequency range $[3,2048]$~Hz. The SNR of all considered events is computed to be consistent with a source distance of 40~Mpc, similar to that of GW170817. 

\subsection{Impact of the RTLNs on the waveform}

GW parameter estimation is affected by systematic errors due to partial knowledge of waveform modelling. This leads to a bias on the estimated parameters, whose relevance will increase for high-SNR events, especially those expected in the 3G era. 

The first step of our analysis is to study the impact of the $6.5$-PN tidal terms on the waveform (Eqs.~\eqref{eq:hatLambdaE} to \eqref{eq:hatGamma}), and the bias they may produce on the measurement of the standard, leading-order tidal term $\tilde{\Lambda}$. In particular, we extend the work done in~\cite{Jimenez-Forteza:2018buh}, which focused only on the impact of the spin-tidal terms $\hat \Lambda$ and $\hat\Sigma$ (Eqs.~\eqref{eq:hatLambdaE} and~\eqref{eq:hatLambdaM}). Here, we also include in the analysis the tidal tail term $\hat{K}$~\eqref{eq:hatK} and the term $\hat{\Gamma}$~\eqref{eq:hatGamma}, which depends on the RTLNs of the binary components.

We compute the probability distributions for the tidal deformability through Eq.~\eqref{eq:match_posterior}. We consider the following waveforms:
\begin{itemize}
    \item $h_\text{$6$-PN}$: TaylorF2 waveform\,\footnote{Since we are comparing waveforms which differ only in the tidal part of the GW phase,  the specific order of the point-particle part is irrelevant for the analysis presented in this section.} truncated at $6$-PN order, including the magnetic TLNs computed for a static fluid;
    \item $h_\text{tail}$: TaylorF2 waveform up to $6$-PN, also including the $6.5$-PN tidal tail term $\hat{K}$;
    \item $h_\text{spin-tidal}$: TaylorF2 waveform up to $6$-PN plus the tidal tail term $\hat{K}$ and spin-tidal terms $\hat \Lambda$ and $\hat\Sigma$;
    \item $h_\text{6.5PN}$: TaylorF2 waveform including, besides $\hat{K}$, $\hat \Lambda$ and $\hat\Sigma$, also the static RTLN term $\hat{\Gamma}$. This case includes \emph{all} static tidal terms up to $6.5$-PN order. 
\end{itemize}

\begin{figure*}[th]
\centering
    \includegraphics[width=\textwidth]{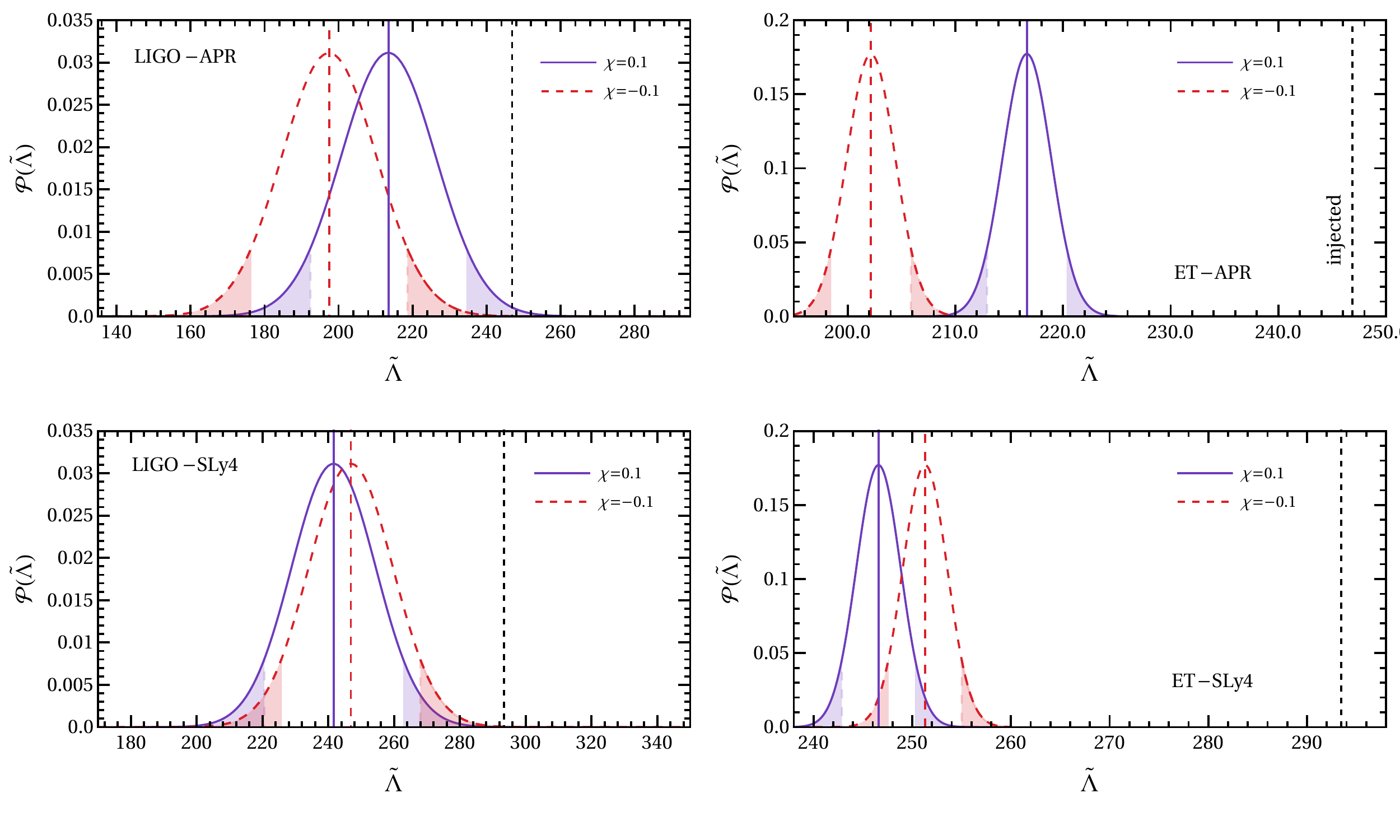}
    \caption{Same as Fig.~\ref{fig:bias_spintidal} but for Setup~II ($h_\text{6-PN}$ vs $h_\text{6.5PN}$), i.e. quantifying the bias introduced by the entire $6.5$-PN term in the waveform. }
    \label{fig:bias}
\end{figure*}

\begin{figure*}[th]
\centering
    \includegraphics[width=\textwidth]{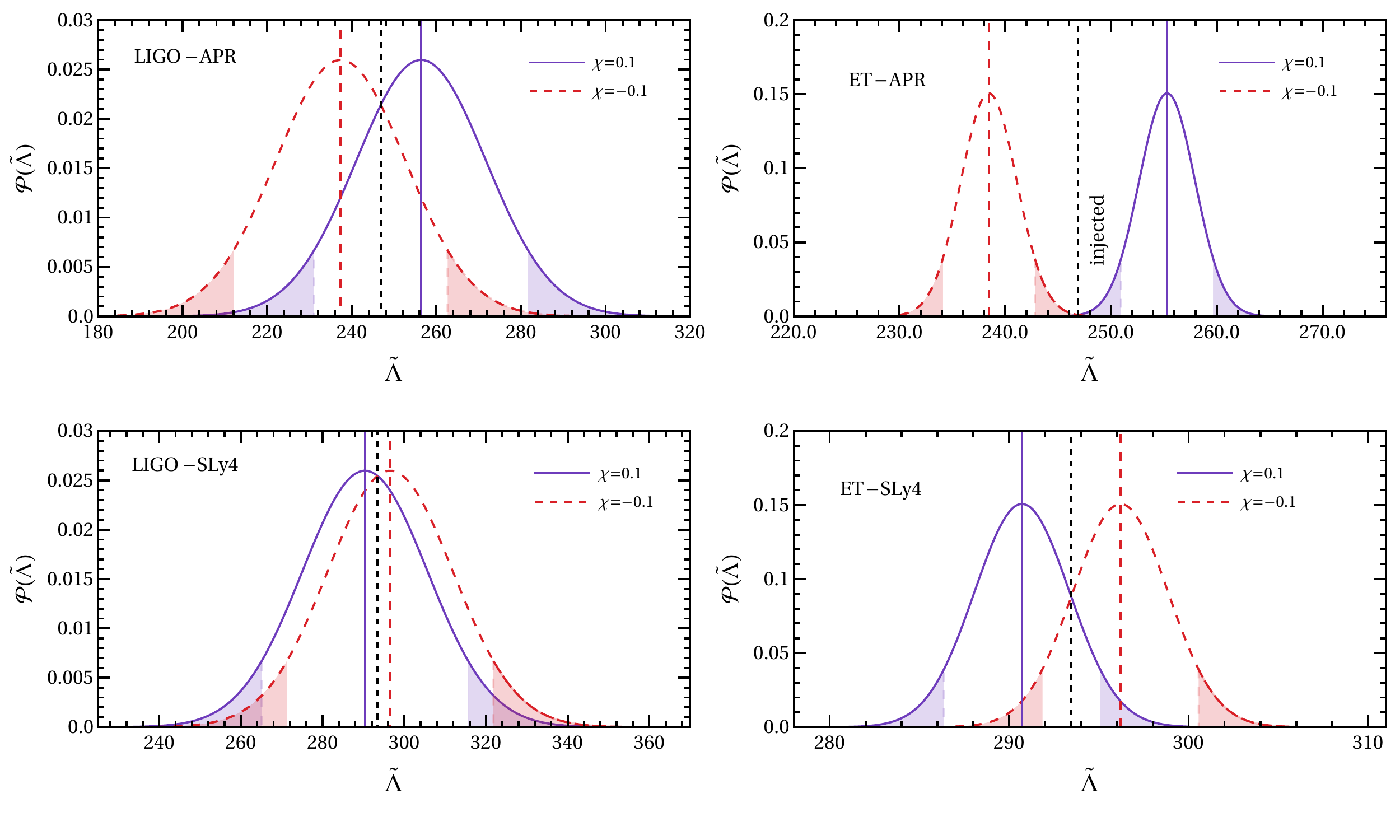}
    \caption{Same as Fig.~\ref{fig:bias} but for Setup~III ($h_\text{6.5PN}$ vs $h_\text{tail}$), i.e., assessing the bias introduced by the tidal tail terms along.
     \label{fig:bias_spin}}
\end{figure*}

This analysis is performed assuming spins $\chi_1=\chi_2=0.1$, for two typical cases of soft and stiff EoS: APR4~\cite{Akmal:1998cf} and SLy4~\cite{Douchin:2001sv}, respectively. For these EoSs we obtain $\Lambda=246,\,293$, respectively, consistent with current LIGO-Virgo observations~\cite{LIGOScientific:2018cki}. These values apply for both NSs, since we assume they are described by the same EoS and have the same mass. Likewise, for those masses we have computed the RTLNs to be $\lambda_{23}=183$, $\lambda_{32}=-247$ for APR4 and $\lambda_{23}=220$, $\lambda_{32}=-159$ for SLy4.
The magnetic TLNs are computed in terms of the electric TLNs via their universal relations~\cite{Yagi:2016bkt}, while the magnetic-led RTLNs are computed through the ``hidden symmetry''~\eqref{eq:hiddens} in terms of the electric-led ones.
The optimal SNR is $\rho=140$ and $\rho=1106$ for the LIGO-Virgo network and for ET, respectively.

We computed the resulting probability distributions in three different setups. Setup I (Fig.~\ref{fig:bias_spintidal}) takes $h_\text{spin-tidal}$ and $h_\text{tail}$ as trigger and template, respectively, replicating Fig.~8 of~\cite{Jimenez-Forteza:2018buh} for realistic equations of state. This shows that, in our current setup, the bias induced in the measurement of $\tilde{\Lambda}$ by the $\hat{\Lambda}$ and $\hat{\Sigma}$ terms alone is  negligible for detections by LIGO-Virgo even at design sensitivity, while it is of the same order of magnitude as the uncertainty for ET detections. 
This motivates the study of 
further contributions at $6.5$-PN order.

Setup II (Fig.~\ref{fig:bias}) takes $h_\text{$6.5$-PN}$ and $h_\text{$6$-PN}$ as trigger and template, respectively, thereby assessing the bias induced by all $6.5$-PN terms in the waveform. We see that the bias is large enough to affect the measurements of $\tilde{\Lambda}$ by both the LIGO-Virgo and ET detectors. 
The largest contributor to this bias is the tidal tail term $\hat{K}$. This can be seen by comparison with Setup~III (Fig.~\ref{fig:bias_spin}), where $h_\text{$6.5$-PN}$ and $h_\text{tail}$ were used as trigger and template, respectively. This includes $\hat{K}$ in the template, hence considering the bias due to the spin-tidal terms alone (including $\hat{\Gamma}$). We see that, while the inclusion of the RTLNs term does not change the irrelevance of the induced bias for LIGO-Virgo detections in comparison with Setup~I, for ET detections the bias can be larger than the expected uncertainty of the measurement of $\tilde{\Lambda}$. However, this result is EoS-dependent: while it occurs for APR4, for SLy4 the spin-TLNs terms and the RTLNs term give opposite sign contributions to the phase of the waveform, canceling each other out.

Our results suggest that, while the largest bias is associated to the tidal tail $\hat K$ (which is effectively included is the most up-to-date waveform templates~\cite{Dietrich:2017aum,Dietrich:2018uni}), \emph{all} $6.5$PN  tidal terms should be included in the modeling of the waveform phase in the data analysis with 3G detectors.

We have further studied the possibility of correlations between different parameters having a significant effect on the expected bias, by varying simultaneously both $\tilde\Lambda$ and $\delta\tilde{\Lambda}$, finding that our conclusions do not change.

\subsection{Measurability of the RTLNs}

Given the prospect of RTLNs having a non-negligible effect on parameter estimation in the 3G era, the natural follow-up question is whether the RTLNs themselves will be measurable. To study this problem we have applied the FIM analysis introduced in Sec.~\ref{sec:stat}, 
computing the probability distribution of the parameters through Eq.~\eqref{eq:fisher_posterior}, and  extracting the associated uncertainty. We did so for the set of parameters
$\{t_c,\phi_c,\log\mathcal{M},\log\nu,\chi_1,\chi_2,\tilde{\Lambda},\delta\tilde{\Lambda},\log\hat{\Gamma}\}$, where $t_c$ and $\phi_c$ are the time and phase at coalescence, $\mathcal{M}$ is the chirp mass, and $\nu$ is the symmetric mass ratio.

We consider binary NS systems with $M_1=1.5M_\odot$, $M_2 = 1M_\odot$; we choose unequal mass binaries (with mass ratio compatible with those of observed systems) because this improves the measurability of the TLNs. We consider NSs described by the APR4 and SLy4 EoSs.
We choose two different values of (small) equal spins $\chi_1=\chi_2=\chi=0.05,0.1$.

As far as the priors are concerned, we assume that spins are normally distributed around zero with variance $\sigma_0(\chi)=1$. For the priors on $\delta\tilde\Lambda$, from Eq.~\eqref{eq:deltaLambda} and our assumption that $M_1>M_2$, we see that $\delta\tilde\Lambda>0$. Thus, we impose a uniform prior ${\cal U}[0,\delta\tilde{\Lambda}_\textnormal{max}]$, with $\delta\tilde{\Lambda}_\textnormal{max}=1000$. We have checked that the upper bound $\delta\tilde{\Lambda}_\textnormal{max}$ is large enough to contain the full posterior, and that increasing it does not affect significantly our results.


\begin{center}
\begin{table*}
\begin{tabular}{c|c|ccccccc}
\hline
\hline
 & EoS & $\mathcal{M}(M_\odot)$ & $\nu$ & $\chi_1$ & $\chi_2$ & $\tilde{\Lambda}$ & $\delta\tilde{\Lambda}$ & $\hat{\Gamma}$ \\ \hline
\multirow{2}{*}{$\chi=0.1$} & APR4 & ${1.06}_{-1.6\times 10^{-6}}^{+1.9\times 10^{-6}}$ & $0.24_{-0.0012}^{+0.0009}$ & $0.1_{-0.15}^{+0.09}$ & $0.1_{-0.15}^{+0.24}$ & $550_{-110}^{+120}$ & $140_{-100}^{+930}$ & $7.7_{-2.2}^{+2.4}\times 10^4$ \\ 
 & SLy4 & ${1.06}_{-1.7\times 10^{-6}}^{+1.9\times 10^{-6}}$ & $0.24_{-0.001}^{+0.001}$ & $0.1_{-0.14}^{+0.10}$
   & $0.1_{-0.15}^{+0.23}$ & $690_{-110}^{+120}$ & $190_{-160}^{+880}$ & $8.4_{-2.3}^{+2.5}\times 10^4$ \\ \hline
\multirow{2}{*}{$\chi=0.05$} & APR4 & ${1.06}_{-1.7\times 10^{-6}}^{+1.7\times 10^{-6}}$ & $0.24_{-0.001}^{+0.001}$ & $0.05_{-0.16}^{+0.13}$ & $0.05_{-0.22}^{+0.26}$ & $550_{-120}^{+120}$ & $140_{-110}^{+930}$ & $3.9_{-2.0}^{+2.6}\times 10^4$\\ 
 & SLy4 & ${1.06}_{-1.7\times 10^{-6}}^{+1.7\times 10^{-6}}$ & $0.24_{-0.001}^{+0.001}$ & $0.05_{-0.15}^{+0.13}$
   & $0.05_{-0.21}^{+0.24}$ & $690_{-110}^{+120}$ & $190_{-160}^{+870}$ & $4.2_{-1.9}^{+2.5}\times 10^4$ \\ 
   \hline
   \hline
\end{tabular}
\caption{Errors on the source parameters assuming an ET detection of a nearly-symmetric NS binary with
two different small values for the spins, and for two tabulated EoS. For each parameter we show the 
the 90\% interval around the median.\label{tab:fisher_spin}}
\end{table*}
\end{center}

The uncertainties computed with this setup are shown in Table~\ref{tab:fisher_spin}. These were extracted from a posterior distribution computed through a Monte Carlo algorithm (see Sec.~\ref{sec:stat}). 

 First of all, we note that the expected uncertainties  on the individual spins are quite large, with the posterior distribution always having support on $\chi=0$ even for $\chi=0.1$. 
We refer the reader to~\cite{inprep2}, where this issue will be studied in detail using a more refined statistical analysis based on Monte Carlo Markov Chain  approaches.

The probability distribution of the RTLN term $\hat\Gamma$ is nearly symmetrical around the injected value, and we find that the relative $90\%$ symmetric confidence interval is approximately $30\%$ for $\chi=0.1$, and $55\%$ for $\chi=0.05$. The errors are larger, of course, for lower values of the spin, or for larger values of the mass ratio (see the discussion below).
Thus, the RTLN term will be  marginally measurable by a 3G detector like ET, for NS spins as large as $\sim0.05-0.1$. 

The relatively large uncertainty of the parameter $\delta\tilde{\Lambda}$ shown in Table~\ref{tab:fisher_spin} motivated us to perform a further analysis, in which we neglect $\delta\tilde{\Lambda}$ as a waveform parameter in the FIM. 
We  find that the uncertainties of all parameters are similar to those obtained with the previous analysis. Therefore,
for the high SNR values expected in the 3G era, $\delta\tilde{\Lambda}$ is predicted to be both unmeasurable and irrelevant for the estimation of the measurement uncertainty of the other parameters, and as such can be neglected in the parameter estimation, although it should be included in the templates for the detection.

\begin{figure}[t]
\centering
    \includegraphics[width=\columnwidth]{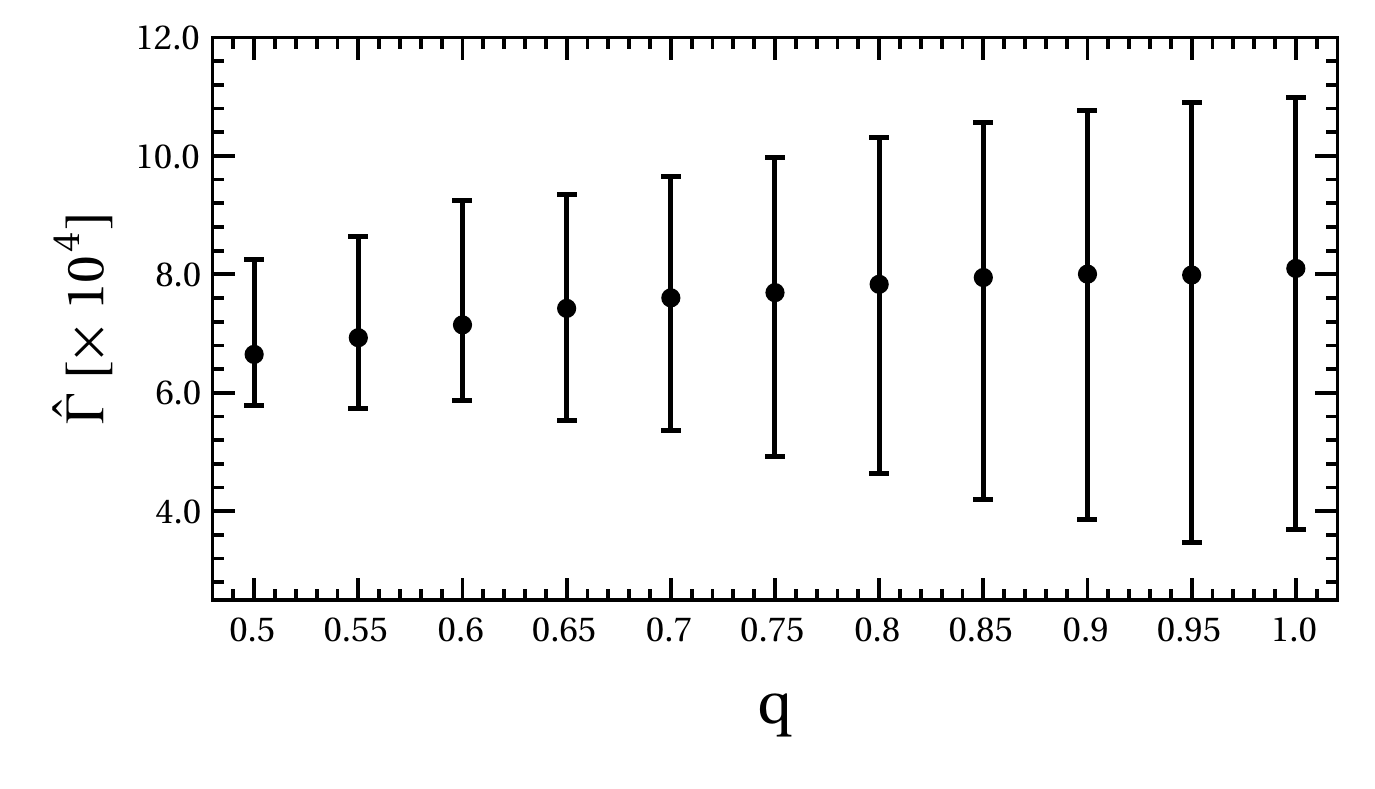}
    \caption{90\% confidence interval around the 
    recovered median value of $\hat{\Gamma}$ as a function of 
    the binary mass ratio $q$.}
    \label{fig:fisher_massratio}
\end{figure}

Finally, we have studied how the binary mass ratio $q=M_2/M_1$ affects the measurability of $\hat{\Gamma}$. We did so for a system with $\chi_1=\chi_2=0.1$, fixed mass $M_2=1\,M_\odot$, and APR4 EoS. We present the associated uncertainties in Fig.~\ref{fig:fisher_massratio}, finding that they mildly decrease for smaller values of  $q$.

From this analysis, we see that there is a realistic prospect of constraining the $6.5$-PN tidal term $\hat{\Gamma}$ with 3G ground-based detectors. We note, however, that this does not necessarily mean that we can estimate the associated (EoS-dependent) RTLNs. The term $\hat{\Gamma}$ 
(explicit in Eq.~\eqref{eq:hatGamma}) depends not only on the RTLNs, but also on the spins $\chi_1$ and $\chi_2$ of the binary components. As discussed above (see also~\cite{inprep2}), these quantities may not be accurately measured even with 3G detectors.
Therefore, measuring  $\hat{\Gamma}$ would not lead to a further constraint on  the EoS, because the measurement errors on the RTLNs coming from the measurement of the $6.5$-PN term would be dominated by the errors on the individual spins.

\section{Conclusions} \label{sec:conclusions}

We have estimated the impact of the $6.5$-PN tidal terms on the GW parameter estimation of the tidal deformability $\tilde{\Lambda}$ for a NS binary system. We have done so for an event consistent with GW170817, considering realistic EoSs (APR4 and SLy4) and an optimal sky orientation. 

We have found that in order to reduce systematic errors, due to incomplete modelling of the tidal effects, in future parameter estimation with 3G detectors all $6.5$-PN tidal terms should be accounted for in the modelling of the signal. In particular, if the NSs have spins as large as $\sim0.05-0.1$ the impact of the RTLNs will be relevant in the 3G era.

We have also found that, for these values of the spins, the RTLN terms in the waveform will be measurable by 3G detectors, even though the measurement of the RTLNs themselves may be impossible  due to the uncertainties in the individual spins. A more detailed study on the measurability of the individual spins with 3G detectors is forthcoming~\cite{inprep2}.

\begin{acknowledgments}
We thank G.~Franciolini and C.~Pacilio for useful discussions.
P.P. acknowledges financial support provided under the European Union's H2020 ERC, Starting Grant agreement no.~DarkGRA--757480. 
This project has received funding from the European Union’s Horizon 2020 research and innovation programme under the Marie Skłodowska-Curie grant agreement No 101007855.
We also acknowledge support under the MIUR PRIN and FARE programmes (GW-NEXT, 
CUP:~B84I20000100001,  2020KR4KN2), and from the Amaldi Research Center funded by the MIUR program "Dipartimento di Eccellenza" (CUP:~B81I18001170001). This work is partially supported by the PRIN Grant 2020KR4KN2 ``String Theory as a bridge between Gauge Theories and Quantum Gravity''.
\end{acknowledgments}

\appendix

\section{PN expansion of the GW phase}
\label{app:Taylor}
We here show the explicit expression of the TaylorF2 waveform. It is:
\begin{equation}
    \tilde{h}(f) = \mathcal{A}f^{-7/6}e^{i \left(2\pi f t_c - \phi_c -\pi/4 + \psi(f)\right)},
\end{equation}
where $\mathcal{A} = \frac{1}{2d}\sqrt{\frac{5\nu M^{5/3}}{6\pi^{4/3}}}$ is the Newtonian amplitude,
$f=\frac{\omega}{\pi}$ the frequency of the wave (i.e., $x=(\pi Mf)^{2/3}/c^2$),
$d$ is the luminosity distance from the source, and $\psi(f)$ is the PN expansion of the GW phase. The GW phase can be written as
\begin{equation}
    \psi = \psi_\text{pp}^\text{$3.5$-PN} + \psi_\text{pp}^\text{spin, $4$-PN} + \psi_{\bar{Q}} + \psi_\text{T}.
\end{equation}
$\psi_\text{pp}^{3.5-\text{PN}}$ is the non-spinning point-particle contribution to the phase up to $3.5$-PN order, given by~\cite{Buonanno:2009zt}
\begin{widetext}
\begin{equation}
    \begin{split}
		\psi_\text{pp}^\text{$3.5$-PN} = & 1 + \left(\frac{3715}{756} + \frac{55}{9}\nu\right) x - 16\pi x^\frac{3}{2} + 10\left(\frac{3058673}{1016064} + \frac{5429}{1008}\nu + \frac{617}{144} \nu^2\right) x^2 \\
		& + \pi \left(\frac{38645}{756} - \frac{65}{9}\nu\right)\left(1 + 3 \log\left(\sqrt{\frac{x}{x_\text{LSO}}}\right)\right) x^\frac{5}{2} \\
		& + \bigg(\frac{11583231236531}{4694215680} - \frac{640}{3}\pi^2 - \frac{6848}{21}e - \frac{6848}{21}\log\left(4\sqrt{x}\right) \\
		& + \left(-\frac{15737765635}{3048192} + \frac{2255}{12}\pi^2\right)\nu + \frac{76055}{1728}\nu^2 - \frac{127825}{1296}\nu^3\bigg) x^3 \\
		& + \pi \left(\frac{77096675}{254016} + \frac{378515}{1512}\nu - \frac{74045}{756}\nu^2\right) x^\frac{7}{2}\,,
	\end{split}
\end{equation}
\end{widetext}
where $x_\text{LSO}=6^{-\frac{3}{2}}$ refers to the frequency at the last stable orbit. $\psi_\text{pp}^{4\text{-PN}}$ is the spinning part of the point-particle contribution to the phase up to $4$-PN order,
\begin{equation}
\begin{split}
	& \psi_\text{pp}^\text{spin, 4-PN} =\, 4\beta_\text{1.5} x^\frac{3}{2} - 10\sigma x^2 + \left[\frac{40}{9}\beta_\text{2.5}\right. \\
	& \left. - \beta_\text{1.5}\left(\frac{3715}{189}+\frac{220}{9}\nu\right)\right]\log\left(x^\frac{3}{2}\right) x^\frac{5}{2}\\
	&+ \mathcal{P}_6 x^3 + \mathcal{P}_7 x^\frac{3}{2} + \mathcal{P}_8 x^4.
\end{split}
\end{equation}
The $1.5$-PN\,\cite{PhysRevD.47.R4183,PhysRevD.52.821,PhysRevD.48.1860}, $2$-PN\,\cite{PhysRevD.71.124043,PhysRevD.57.5287} and $2.5$-PN\,\cite{PhysRevD.74.104034} coefficients, assuming aligned spins, are given by
\begin{equation}
	\beta_\text{1.5} = \left(\frac{113}{12}\eta_1^2 + \frac{25}{4}\nu\right)\chi_1 + \left(\frac{113}{12}\eta_2^2 + \frac{25}{4}\nu\right)\chi_2,
\end{equation}
\begin{equation}
	\sigma = \frac{79}{8}\nu\chi_1\chi_2 + \frac{81}{16}\left(\eta_1^2\chi_1^2+\eta_2^2\chi_2^2\right),
\end{equation}
\begin{equation}
	\begin{split}
	    \beta_\text{2.5} = & \left[\left(-\frac{31319}{1008} + \frac{1159}{24}\nu\right)\eta_1^2 + \left(-\frac{809}{84}+\frac{281}{8}\nu\right)\right]\chi_1 \\
	    & +(1\leftrightarrow 2).
	\end{split}
\end{equation}
The subsequent $3$-PN, $3.5$-PN and $4$-PN order terms can be found in\,\cite{Mishra:2016whh}
\begin{widetext}
\begin{align}	
	\mathcal{P}_6 = &\pi \left(\frac{2270}{3}\delta\chi_a + \left(\frac{2270}{3}-520\nu\right)\chi_s\right) + \left(\frac{75515}{144} - \frac{8225}{18}\nu\right)\delta \chi_a \chi_s \\
	& + \left(\frac{75515}{288} - \frac{263245}{252}\nu - 480\nu^2\right)\chi_a^2 + \left(\frac{75515}{288} - \frac{232415}{504}\nu + \frac{1255}{9}\nu^2\right)\chi_s^2, \\
	\mathcal{P}_7 = & \left(-\frac{25150083775}{3048192} + \frac{26804935}{6048}\nu - \frac{1985}{48}\nu^2\right)\delta \chi_a \\
	& + \left(-\frac{25150083775}{3048192} + \frac{10566655595}{762048}\nu - \frac{1042165}{3024}\nu^2 + \frac{5345}{36}\nu^3\right)\chi_s + \left(\frac{14585}{24}-2380\nu\right)\delta\chi_a^3 \\
	& + \left(\frac{14585}{24} - \frac{475}{6}\nu + \frac{100}{3}\nu^2\right)\chi_s^3 + \left(\frac{14585}{8} - \frac{215}{2}\nu\right)\delta \chi_a\chi_s^2 + \left(\frac{14585}{8} - 7270\nu + 80\nu^2\right)\chi_a^2\chi_s, \\
	\mathcal{P}_8 & = \pi\left[\left(\frac{233915}{168} - \frac{99185}{252}\nu\right)\delta \chi_a + \left(\frac{233915}{168} - \frac{3970375}{2268}\nu + \frac{19655}{189}\nu^2\right)\chi_s\right]\left(1-3\log\left(\sqrt{x}\right)\right),
\end{align}
\end{widetext}
with $\delta=\eta_1-\eta_2$, $\chi_s = \frac{\chi_1+\chi_2}{2}$, $\chi_a = \frac{\chi_1-\chi_2}{2}$ defined for the sake of readability. $\psi_{\bar{Q}}$ is the correction of the point-particle phase to include the quadrupole moment of a NS,
\begin{equation}
    \begin{split}
      \psi_{\bar{Q}} = & -50\left((\eta_1^2\chi_1^2+\eta_2^2\chi_2^2)(\bar{Q}_s-1)\right. \\
       & \left. + (\eta_1^2\chi_1^2-\eta_2^2\chi_2^2)\bar{Q}_a\right),
    \end{split}
\end{equation}
where $\bar{Q}_s = \frac{\bar{Q}_1 + \bar{Q}_2}{2}$ and $\bar{Q}_a = \frac{\bar{Q}_1 - \bar{Q}_2}{2}$, 
with $\bar{Q}_A=\frac{Q_A}{M_A^3\chi_A^2}$ normalized mass quadrupole of body $A$.
In this article, the quadrupole moment has been  computed through the I-Love-Q relations derived in \cite{Yagi:2013awa}.
Finally, the tidal phase $\psi_T$ is defined in Eq.~\eqref{eq:psiT}.
\bibliographystyle{apsrev4-1}
\bibliography{biblio}

\end{document}